\documentclass[aps,prl,reprint,superscriptaddress,showpacs]{revtex4-1}

\usepackage{graphicx}
\usepackage{color}
\usepackage{booktabs}

\begin{document}


\title{Frequency ratio of two optical clock transitions in ${}^{171}$Yb$^{+}$ and constraints on the time-variation of fundamental constants}


\author{R.\,M.\,Godun}
\email[]{rachel.godun@npl.co.uk}
\author{P.\,B.\,R.\,Nisbet-Jones}
\affiliation{National Physical Laboratory, Hampton Road, Teddington, TW11 0LW, UK}
\author{J.\,M.\,Jones}
\affiliation{National Physical Laboratory, Hampton Road, Teddington, TW11 0LW, UK}
\affiliation{School of Physics and Astronomy, University of Birmingham, Birmingham, B15 2TT, UK}
\author{S.\,A.\,King}
\author{L.\,A.\,M.\,Johnson}
\author{H.\,S.\,Margolis}
\author{K.\,Szymaniec}
\author{S.\,N.\,Lea}
\affiliation{National Physical Laboratory, Hampton Road, Teddington, TW11 0LW, UK}
\author{K.\,Bongs}
\affiliation{School of Physics and Astronomy, University of Birmingham, Birmingham, B15 2TT, UK}
\author{P.\,Gill}
\affiliation{National Physical Laboratory, Hampton Road, Teddington, TW11 0LW, UK}




\begin{abstract}
Singly-ionized ytterbium, with ultra-narrow optical clock transitions at 467~nm and 436~nm, is a convenient system for the realization of optical atomic clocks and tests of present-day variation of fundamental constants.  We present the first direct measurement of the frequency ratio of these two clock transitions, without reference to a cesium primary standard, and using the same single ion of ${}^{171}$Yb${}^{+}$.  The absolute frequencies of both transitions are also presented, each with a relative standard uncertainty of $6\times 10^{-16}$.  Combining our results with those from other experiments, we report a three-fold improvement in the constraint on the time-variation of the proton-to-electron mass ratio, $\dot{\mu}/{\mu} = 0.2(1.1)\times 10^{-16}\,{\rm year}^{-1}$, along with an improved constraint on time-variation of the fine structure constant, $\dot{\alpha}/{\alpha} = -0.7(2.1)\times 10^{-17}\,{\rm year}^{-1}$.
\end{abstract}


\maketitle


A redefinition of the SI second is expected to be based on optical frequency standards since their performance, in terms of both statistical and systematic uncertainties~\cite{Bloom:BestSrLatt,Hinkley:YbLattStab,LeTargat:Sr,Chou:Al,Barwood:Sr+twotrap,Madej:Sr+,Rosenband:Alpha}, can exceed that of cesium primary standards~\cite{Guena:Fountain,Heavner:Fountain}.  Presently, singly-ionized ytterbium is unique amongst optical frequency standards in that it possesses two transitions that are accepted as secondary representations of the second: an electric quadrupole (E2: ${}^{2}$S${}_{1/2}\rightarrow{}^{2}$D${}_{3/2}$) transition at 436~nm~\cite{Webster:Quadrupole,Tamm:OpticalCs} and an electric octupole (E3: ${}^{2}$S${}_{1/2}\rightarrow{}^{2}$F${}_{7/2}$) transition at 467~nm~\cite{King:Octupole,Huntemann:Octupole}. These transitions have an exceptionally large differential sensitivity to time-variation of the fine structure constant, which allows important tests of fundamental physics.  The absolute frequencies of both the E3 and E2 transitions have been measured with increasingly high accuracy and are now limited by Cs primary standards.  A direct measurement of the ratio of the two optical frequencies is, however, free from the additional uncertainties introduced by the primary standard.  Direct optical ratio measurements will therefore allow international comparisons of optical frequency standards to be made with lower statistical and systematic uncertainties than is currently achievable through absolute measurements, an essential task prior to a redefinition of the SI second.

In this Letter, we present the first direct measurement of the ratio between the E3 and E2 optical transition frequencies in ${}^{171}$Yb${}^{+}$, with fractional uncertainty $3\times 10^{-16}$.  New absolute frequency measurements of both transitions are also presented, with fractional uncertainty $6\times 10^{-16}$.  These measurements set new constraints on present-day time-variation of the fine structure constant, $\alpha$, and the proton-to-electron mass ratio, $\mu$.

\begin{figure}
\includegraphics[width=3.5in]{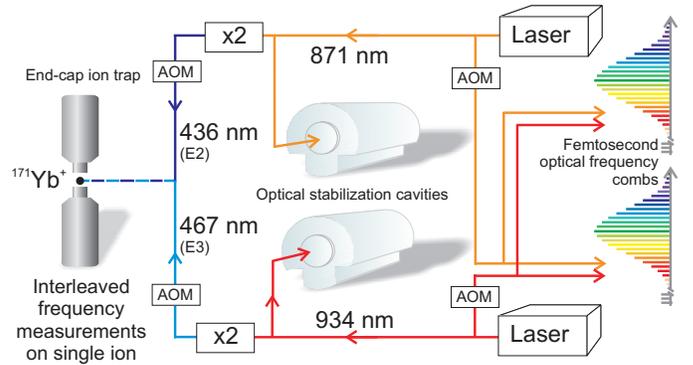}
\caption{\label{fig_1}(color online). Schematic experimental arrangement for measuring two clock transition frequencies simultaneously in a single ion, as described in the text.}
\end{figure}

Detailed descriptions of the NPL ytterbium ion optical frequency standard can be found in references~\cite{Webster:Quadrupole,King:Octupole}. The atomic reference is provided by a single ion of ${}^{171}$Yb${}^{+}$, trapped and laser-cooled in an end-cap trap~\cite{Schrama:Endcap}. The frequency ratio $\nu_{\rm{E3}}/\nu_{\rm{E2}}$ is determined by stabilizing lasers to the E3 and E2 transitions and measuring the ratio between the laser frequencies with femtosecond optical frequency combs, as depicted in Fig.~\ref{fig_1}. The E3 and E2 ultrastable lasers are frequency-doubled extended-cavity diode lasers, stabilized to independent high-finesse cavities.  The ion is excited on the $m_F=0 \rightarrow m_F'=0$ component of each clock transition by a rectangular pulse of length $\tau$ from the appropriate ultrastable laser ($\tau=100$~ms and $30$~ms for E3 and E2 respectively) with the success of the excitation determined by the electron-shelving technique.  Each laser is stabilized to its atomic transition by probing at the estimated half-maxima on both the high and low frequency sides of the Fourier-limited lineshape and generating an error signal based on the imbalance between the two quantum jump rates. Feedback is applied by computer control of the drive frequency to acousto-optic modulators (AOMs) in the blue beam paths. The AOM frequencies in the infrared beam paths are also updated to maintain a fixed relationship between the clock transition frequencies and the laser frequencies sent to the combs \cite{King:Octupole}. These AOMs are also used to phase-stabilize the 50~m optical fibre links between the ultrastable lasers and the combs. By interleaving the probe pulses on a $\sim$100 ms timescale, both lasers are simultaneously stabilized to their respective transitions in the same ion. This brings experimental simplicity and common-mode rejection of certain systematic effects such as the gravitational redshift and relativistic time dilation.

\begin{table*}
\begin{ruledtabular}
\begin{tabular}{lrrrrr}
& \multicolumn{2}{c}{\textbf{E2 transition}} & \multicolumn{2}{c}{\textbf{E3 transition}} & \textbf{Ratio E3/E2}\\
\textbf{Source of shift}             & $\delta\nu/\nu_0 (10^{-16})$ & $\sigma/\nu_0 (10^{-16})$ & $\delta\nu/\nu_0 (10^{-16})$  & $\sigma/\nu_0 (10^{-16})$    &$\sigma/{\rm ratio} (10^{-16})$\\ \hline
Residual quadrupole         & 0         	   & 2.91              & 0         	       & 0.06  		          & 2.97      		\\
BBR: Polarizability         &-4.86     	       & 0.99      		   &-0.98     	       & 0.45                 & 1.09      		\\
BBR: Trap temperature    	& 0	               & 0.13      		   & 0         	       & 0.03  		          & 0.11      		\\
Quadratic Zeeman (static + ac) & 75.76     	   & 0.77      		   &-3.24     	       & 0.04  		          & 0.80      		\\
Quadratic rf Stark       	&-0.41     	       & 0.41      		   & -0.09     	       & 0.09  		          & 0.33      		\\
Residual ac Stark (probe laser)& 0             & 0.02      	       & 0                 & 0.20  		          & 0.20      		\\
Servo error                 & 0         	   & 0.07      		   & 0         	       & 0.08  		          & 0.11      		\\
AOM phase chirp             & 0         	   & 0.06      		   & 0         	       & 0.06  		          & 0.09      		\\
ac Stark (370, 760 and 935 nm) & 0       & 0.07      	       & 0         	       & 0.08  		          & $<$0.01   		\\ \hline
\textbf{SUBTOTAL}           & \textbf{70.49}   &\textbf{3.19}      &\textbf{-4.31}     &\textbf{0.52}         &\textbf{3.29}    \\ \hline
Additional (TABLE II)       & -0.81   	       & 2.16  		       & -0.81   	       & 2.16  		          & - 			    \\
Statistical                 & -       	       & 4.77	    	   & -       	       & 5.35 			      & 0.68 			\\ \hline
\textbf{TOTAL}              &\textbf{69.68}    &\textbf{6.13}      &\textbf{-5.12}     &\textbf{5.79}         &\textbf{3.36}    \\
\end{tabular}
\end{ruledtabular}
\caption{Summary of the leading systematic shifts and associated relative standard uncertainties relevant to the absolute measurements and the optical frequency ratio measurement.}
\label{Tab:ratiosystematics}
\end{table*}

\begin{table}
\begin{ruledtabular}
\begin{tabular}{lrr}
& \multicolumn{2}{c}{\textbf{E2 \& E3  transition}} \\
\textbf{Source}             & $\delta\nu/\nu_0 (10^{-16})$ & $\sigma/\nu_0 (10^{-16})$ \\ \hline
Gravitational redshift      & -0.71     & 0.16\\
Second-order Doppler        & -0.10     & 0.10\\
Cs fountain systematics     & 0         & 1.90\\
Comb, rf distribution \& synthesis         & 0         & 1.00\\ \hline
\textbf{SUBTOTAL}           & \textbf{-0.81}  & \textbf{2.16} \\
\end{tabular}
\end{ruledtabular}
\caption{Summary of the additional systematic shifts common to the absolute frequency measurement of both transitions, with their associated relative standard uncertainties.}
\label{Tab:quadsystematics}
\end{table}

The systematic shifts and associated uncertainties are listed in Table~\ref{Tab:ratiosystematics}.  In some cases, such as the blackbody shift arising from the trap temperature, the systematic frequency shifts for the two transitions are correlated and so the uncertainty for the ratio differs from the quadrature sum of the two.  The largest systematic shift in this experiment ($\sim$600 Hz) was the ac Stark shift of the E3 transition, arising from the relatively high intensity ($\sim$300~W/cm${}^2$) required to drive the nanohertz linewidth transition at 467~nm~\cite{Roberts:171Octupole}.  The laser power was stabilized by active feedback to the drive amplitude of the AOM in the 467~nm beam path.  The ac Stark shift was measured and extrapolated to the unperturbed frequency $\nu_0$ by probing the transition at two different laser power levels, $P_1$ and $P_2$. Assuming constant beam pointing over the $\sim$3~s cycle time, and with accurate knowledge of the power ratio $\kappa=P_2/P_1$, $\nu_0$ can be found using the relation $\nu_{0} = (\kappa\nu_1-\nu_2)/(\kappa -1)$, where $\nu_i$ is the perturbed frequency when probing with power $P_i$.  Data was taken with power ratios of both 5:1 and 3:1 in order to vary the systematic conditions and check the reliability of the extrapolation procedure. The ratio $\kappa$ was monitored continuously using a photodiode with calibrated linearity. A fractional uncertainty of $\sigma_{\kappa}/\kappa = 2\times 10^{-4}$ gave the leading contribution to the fractional uncertainty of the ac Stark shift at a level of $1.8\times 10^{-17}$. A smaller contribution to the ac Stark shift uncertainty came from the laser power servo for each probe pulse taking $\sim 200~\mu$s to settle at the start of the pulse.  During this time (about 0.2\% of the pulse duration), the ac Stark shift of the atomic transition differed from its static value.  Measurement of the temporal profile of the probe pulse during servo capture, combined with numerical modeling, predicted a contribution to the fractional uncertainty of the ac Stark shift at a level of $8\times 10^{-18}$.

The ac Stark shift was removed in real-time during the measurement of the optical frequency ratio by interleaving three simultaneous, independent servos in order to stabilize (1)~the 436~nm laser frequency, as well as (2)~the high power and (3)~the low power laser frequencies to the 467~nm perturbed transitions.  At the end of each 3~s servo cycle, when all three transitions had been probed four times on each of the low and high frequency sides of their respective line centers, all AOM center frequencies were simultaneously updated for the new E2 and ac Stark-free E3 frequencies.

The largest contribution to the uncertainty in the frequency ratio comes from the quadrupole shift of the E2 transition. This arises from the interaction between the quadrupole moment of the upper state of the clock transition and any stray electric field gradient within the trap. The magnitude of the shift, $\Delta\nu\propto 3\cos^{2}\theta -1$, is a function of $\theta$, the relative angle between the field gradient and the quantization axis.  This shift averages to zero when the transition is interrogated in three mutually orthogonal magnetic field orientations~\cite{Itano:Quadrupole}. The field direction was therefore switched every 5~minutes in our experiments so that the frequency obtained after fifteen minute rolling averages was free of the quadrupole shift. There is, however, an uncertainty in the residual frequency shift due to the uncertainty on the orthogonality of the magnetic field directions.  The magnitudes of the three fields used for data taking were monitored before and after every data set, along with a second set of nominally orthogonal fields. Knowledge of field magnitudes in six directions greatly constrained the uncertainty on non-orthogonality in the set fields. The field magnitudes, evaluated from first-order Zeeman splittings of the E2 transition, were all $10.00(5)~\mu$T.  For the ion trap used in this experiment, residual electric field gradients resulted in tensorial quadrupole shifts of $+0.5$~Hz, $+9.5$~Hz, and $-10$~Hz on the E2 transition for the three selected operating fields.  Simulations of the quadrupole shift nulling under the above magnetic field constraints were performed, which led to fractional uncertainties of $2.9\times 10^{-16}$ for the E2 transition.

The blackbody shift, which is the dominant systematic uncertainty for the E3 transition, was determined using experimental values for the differential polarizabilities of the atomic states for each transition~\cite{Huntemann:Octupole,Schneider:QuadMoment}. The ion experienced a trap temperature of $294.5 \pm 2$~K, which was verified through theoretical modeling and IR measurement of an identical dummy trap \cite{CMI:Temp}.  The shift uncertainties are dominated by the uncertainty in the polarizability coefficients; the contribution from temperature uncertainty alone would give rise to fractional uncertainties of just $\sigma_{\rm{E2}}/\nu_{\rm{E2}} =1.3\times10^{-17}$ and $\sigma_{\rm{E3}}/\nu_{\rm{E3}} =2.6\times10^{-18}$.

\begin{figure}
\includegraphics[width=3.5in]{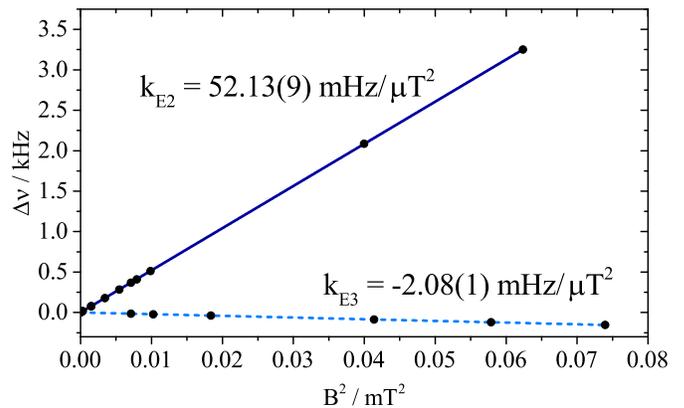}
\caption{\label{fig_2}(color online). Quadratic Zeeman shifts, $\Delta \nu$ of the $m_F=0 \rightarrow m_F=0$ component, plotted for the E2 (solid line) and E3 (dashed line) transitions as a function of magnetic field.  The error bars on each point are too small to show.}
\end{figure}

The second-order Zeeman shifts of the clock transitions, $\Delta \nu_i = k_i B^2$, were evaluated from measurements of the applied magnetic field $B$, combined with the appropriate shift coefficient $k_{i}$.  In the published literature, for $k_{\rm{E2}}$ there is only a calculated value~\cite{Tamm:OpticalCs}, and for $k_{\rm{E3}}$ the experimental values~\cite{Webster:Zeeman,Hosaka:Zeeman} were determined in a less direct way than could be measured in our current setup.  We therefore made new measurements of the coefficients by changing the magnetic field (with magnitude determined from Zeeman component splittings of the E2 transition), and observed the change in the clock transition frequency relative to a reference with constant magnetic field. The results are plotted in Fig.~\ref{fig_2} and lead to measured values of $k_{\rm{E2}}=52.13(9)$~mHz/$\mu$T${}^{2}$ and $k_{\rm{E3}}=-2.08(1)$~mHz/$\mu$T${}^{2}$.  This is the first directly measured experimental value for $k_{\rm{E2}}$ and it agrees well with theory.

\begin{figure}
\includegraphics[width=3.5in]{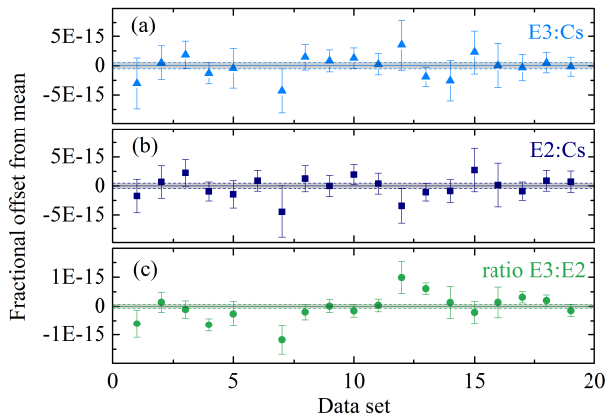}
\caption{\label{fig_3}(color online). (a,b) Absolute frequencies and (c) ratios measured in February and March 2014. The solid lines show the weighted means of the data, which have uncertainties given by the dashed lines. The error bars on the individual points indicate their statistical uncertainty.}
\end{figure}

The optical frequency ratio was measured simultaneously on two independent femtosecond frequency combs~\cite{Walton:Comb,Tsatourian:Comb}, both using the transfer-oscillator scheme~\cite{Telle:TransferOscillator}, which removes the requirement for tight locking of the comb modes to an optical standard.  For the data presented in this paper, the instability of the measured frequency ratio was $\sim3.6\times10^{-14}~\tau^{-1/2}$ for averaging times $\tau$ grater than the 15-minute measurement cycle, and the agreement between the two combs was $\sim$1$\times10^{-20}$ \cite{Johnson:Comb}.    After correction for all systematic shifts, the ratio was determined to be $\nu_{\rm{E3}}/\nu_{\rm{E2}} = 0.932~829~404~530~964~65(31)$. The results are shown in Fig.~\ref{fig_3}(c) with the error bars representing statistical uncertainties only.  The cesium primary frequency standard NPL-CsF2~\cite{Li:Fountain,Szymaniec:Fountain2014} was also operating during the measurement period, enabling absolute frequencies to be recorded simultaneously, and these are also shown in Fig.~\ref{fig_3}.  The E2 and E3 optical frequencies were recorded for 105 and 81 hours respectively, with 72 hours of overlap. Extrapolating the experimentally observed instabilities to the total measurement period leads to a statistical relative standard uncertainty on the mean ratio of $7\times10^{-17}$, and $5\times10^{-16}$ on each of the absolute frequencies. Although the statistical uncertainties for the absolute frequencies are much higher than for the ratio, the absolute measurements still provide an important self-consistency check. The E3 absolute frequency reported here has lower uncertainty than the best previously published value~\cite{Huntemann:Octupole}.  The final results $\nu_{\rm{E3}}$ = 642~121~496~772~644.91(37)~Hz and $\nu_{\rm{E2}}$ = 688~358~979~309~308.42(42)~Hz agree well with previous measurements.  Furthermore, the excellent agreement with previously published values from PTB~\cite{Huntemann:Octupole,Tamm:OpticalCs}, leads to the E3 and E2 transitions in $^{171}$Yb$^+$ now having the best and second best international agreement for trapped-ion optical frequency standards respectively.

The history of absolute frequency measurements (AFM) in ${}^{171}$Yb${}^{+}$ and other frequency standards worldwide
can provide a constraint on present-day time-variation of the fine structure constant, $\alpha$,
and the proton-to-electron mass ratio, $\mu = m_{\rm p}/m_{\rm e}$,
on the assumption of local position invariance and that any time-variation is linear. Variation of the fundamental constants is predicted by some grand unified theories that go beyond the standard model~\cite{Marciano:GUT,Damour:GUT,Uzan:GUT}. Measurements of radiation emitted from atoms in distant galaxies $\sim$10$^{10}$~years ago have provided conflicting results, with claims of time variation \cite{Webb:FurtherAlphadot,Webb:SpatialAlphadot} and null results \cite{Quast:AlphadotNull,Bagdonaite:mudotMeth} both reported.  Laboratory experiments offer complementary tests of present-day variation.

Time variation of $\alpha$  and $\mu$ is measured via the frequency ratio of two clock transitions $r=\nu_1/\nu_2$ using the relation $\dot{r}/r = (A_1-A_2)\dot{\alpha}/\alpha + (B_1-B_2)\dot{\mu}/\mu$. The sensitivity coefficients $A_i$ of the relevant atomic transitions are: $A_{\rm{E3}} = -5.95$~\cite{Dzuba:Octupole}, $A_{\rm{E2}} = 0.88$~\cite{Dzuba:Quadrupole}, and  $A_{\rm{Cs}} = 2.83$~\cite{Dzuba:CS}. Dependence on $\dot{\mu}$ arises through the nuclear magnetic moment and is negligible for optical transitions $($the sensitivity coefficient $B_{\rm{E3}}=B_{\rm{E2}}=0)$. As the $^{133}$Cs nuclear {\it g}\,-factor has negligible sensitivity to changes in the strong interaction~\cite{Flambaum:Strong,Berengut:Constants},  an AFM history can also be interpreted to set constraints on $\dot{\mu}/\mu$ ($B_{\rm{Cs}}=-1$).

The present E3 and E2 absolute frequency results, in combination with previous measurements at
NPL~\cite{Hosaka:Octupole,King:Octupole,Webster:Quadrupole} and at PTB~\cite{Huntemann:Octupole,Stenger:Quadrupole,Peik:Alphadot,Tamm:Quadrupole07,Tamm:QuadShift,Tamm:OpticalCs},
place a limit on linear time-variation of the fine structure constant at the level of
$[\dot{\alpha}/{\alpha}]_{\rm Yb^+/Cs} = 7.2(4.7)\times 10^{-17}\,{\rm year}^{-1}$   yielding the constraint $[\dot{\mu}/{\mu}]_{\rm Yb^+/Cs} = 3.5(2.4)\times 10^{-16}\,\rm{year}^{-1}$.

\begin{figure}
\includegraphics[width=3.5in]{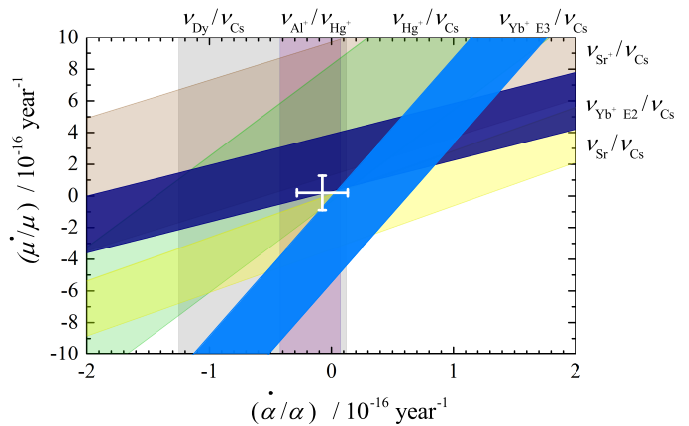}
\caption{\label{fig_4}(color online). The shaded regions represent the constraints placed on time-variation of the fine structure constant, $\alpha$ and the proton-to-electron mass ratio, $\mu$, from a history of frequency measurements in ${}^{171}$Yb$^{+}$, ${}^{88}$Sr$^{+}$ \cite{Margolis:Sr_ion, Madej:Sr+, Barwood:Sr+twotrap}, ${}^{87}$Sr \cite{LeTargat:Sr, Falke:Sr} ${}^{199}$Hg$^{+}$\cite{Fortier:Hg_ion}, Dy \cite{Leefer:Dysprosium} and ${}^{27}$Al$^{+}$  \cite{Rosenband:Alpha}.  The white error bars represent the standard uncertainty in these values when combining the constraints from all transitions.}
\end{figure}

Including data from other experiments in the analysis can further improve these constraints (Figure~\ref{fig_4}). Due to the high sensitivity coefficients of the E3 and E2 transitions in ${}^{171}$Yb${}^{+}$, inclusion of data from the other AFMs only marginally improves upon the Yb$^{+}$/Cs limits, to give $[\dot{\alpha}/{\alpha}]_{\rm AFM} = 4.4(4.4)\times 10^{-17}\,{\rm year}^{-1}$ and $[\dot{\mu}/{\mu}]_{\rm AFM} = 1.4(1.9)\times 10^{-16}\,{\rm year}^{-1}$.  Limits have also been reported from an experiment in atomic dysprosium~\cite{Leefer:Dysprosium} giving $[\dot{\alpha}/{\alpha}]_{\rm Dy} = -5.8(6.9)\times 10^{-17}\,{\rm year}^{-1}$, and from an optical frequency ratio of transitions in Al$^+$ and Hg$^+$~\cite{Rosenband:Alpha} giving $[\dot{\alpha}/{\alpha}]_{\rm Al^+/Hg^+} = -1.7(2.5)\times 10^{-17}\,{\rm year}^{-1}$ (using a revised value of the sensitivity coefficient for the transition in Hg$^+$~\cite{Dzuba:Octupole}).  Combining these independent values gives a new limit to present-day time-variation of the fine structure constant, $\dot{\alpha}/{\alpha} = -0.7(2.1)\times 10^{-17}\,{\rm year}^{-1}$, also leading to a new limit on time-variation of the proton-to-electron mass ratio, $\dot{\mu}/{\mu} = 0.2(1.1)\times 10^{-16}\,{\rm year}^{-1}$.  This is a three-fold improvement on the best previously reported present-day constraint on $\dot{\mu}/{\mu}$~\cite{Guena:mudot}.

Further $\nu_{\rm E3}/\nu_{\rm E2}$ direct optical frequency ratio measurements, free from the additional uncertainties imposed by the Cs standard,
will yield an even tighter constraint on time-variation of $\alpha$~\cite{Lea:Constants}.  Repeated measurements will enable limits to be placed on $\dot{\alpha}/{\alpha}$ at a level below $1 \times 10^{-17}\,{\rm year}^{-1}$ in the next few years.

In conclusion, we have made the first direct measurement of the optical frequency ratio of the E3 and E2 transitions in ${}^{171}$Yb$^{+}$. This value can be used to compare similar optical standards in different institutions, free from the limitations imposed by referencing to local microwave standards. The absolute frequencies of these transitions, relative to a Cs fountain primary standard, were also presented.  The E3 absolute frequency has the lowest uncertainty reported to date for this transition and contributes strongly to revised constraints on present-day time-variation of the fine structure constant and the proton-to-electron mass ratio. We plan to improve the ${}^{171}$Yb${}^{+}$ optical frequency standard with several experimental upgrades, including a hyper-Ramsey interrogation scheme that heavily suppresses the ac Stark shift on the E3 transition \cite{Yudin:HyperRamsey,Huntemann:Hyper}, and a reduction of the quadrupole shift through better cancellation of stray electric field gradients.  Improved measurements of the differential polarizabilities for both the E3 and E2 transitions will further reduce the uncertainty budgets.

\begin{acknowledgments}
This work was funded by the UK National Measurement System, the European Space Agency (ESA) and the European Metrology Research Programme (EMRP). The EMRP is jointly funded by the EMRP participating countries within EURAMET and the European Union. The authors would also like to thank Teresa Ferreiro for assistance with the femtosecond combs.
\end{acknowledgments}

After submission of this manuscript, closely related, but independent, measurements were also presented by PTB~\cite{Huntemann:mudot}.

\bibliography{GodunReferences}
\end{document}